\documentclass[preprint2]{aastex}

\usepackage{graphicx}

\newcommand{\tensor}[1]{\mathsf{#1}}
\newcommand{\text}[1]{\mbox{#1}}
\newcommand{\agt}{\ga}
\newcommand{\alt}{\la}

\newcommand{\grad}[1]{\nabla #1}
\renewcommand{\div}[1]{\nabla \cdot #1}
\newcommand{\curl}[1]{\nabla \times #1}

\newcommand{\B}{\vec{B}}
\renewcommand{\b}{\hat{b}}
\newcommand{\W}{\tensor{W}}
\newcommand{\E}{\vec{E}}
\newcommand{\J}{\vec{J}}
\renewcommand{\v}{\vec{v}}
\newcommand{\z}{\hat{z}}

\begin{document}

\title{Finite Larmor Radius Effects on the Magnetorotational
  Instability}
\author{Nathaniel M. Ferraro} 
\affil{Princeton Plasma Physics Laboratory, Princeton, NJ, 08543-0451}

\begin{abstract}
  The linear dispersion relation for the magnetorotational instability
  (MRI) is derived including finite Larmor radius (FLR) effects.  In
  particular, the Braginskii form of the ion gyroviscosity, which
  represents the first-order FLR corrections to the two-fluid
  equations, is retained.  It is shown that FLR effects are the most
  important effects in the limit of weak magnetic fields, and are much
  more important than the Hall effect when $\beta_i \gg 1$, where
  $\beta_i$ is the ratio of the ion thermal pressure to the magnetic
  pressure.  FLR effects may completely stabilize even MRI modes
  having wavelengths much greater than the ion Larmor radius.  Some
  implications for astrophysical accretion disks are discussed.
\end{abstract}


\maketitle 

\section{Introduction}

The magnetorotational instability is a local instability which may be
present in accretion disks having sheared azimuthal flow and a weak
magnetic field \citep{Balbus91}. Turbulence resulting from this
instability is thought to play an important role in the radial
transport of angular momentum in such systems \citep{Balbus98}.  Here
we explore the modifications to the local, two-fluid theory of the MRI
in the linear regime due to finite Larmor radius (FLR) effects.

\cite{Rosenbluth62} have shown using kinetic theory that FLR effects
can be stabilizing to ``weakly unstable'' modes---defined as modes
having a linear growth rate much smaller than the ion cyclotron
frequency---even when the mode wavelength is much larger than the ion
Larmor radius.  It was later shown by \cite{Roberts62} that this
result could be obtained from fluid theory by retaining the
gyroviscous stress component of the ion pressure tensor.  In most
physical scenarios the MRI is weakly unstable in the sense of
Rosenbluth \textit{et al.}, and indeed we show that the MRI may be
completely stabilized by gyroviscous effects at scales much larger
than the ion Larmor radius.  In some cases this stabilization
significantly constrains the spectrum of linearly unstable modes.

The effect of the Hall term, which accounts for differences between
the electron and ion fluid velocities, has been examined previously by
\cite{Wardle99}, \cite{Balbus01}, \cite{Salmeron03}, and
\cite{Krolik06}.  In particular, it was found that the Hall effect may
be either stabilizing or destabilizing, depending on whether the
equilibrium magnetic field is aligned or anti-aligned to the
equilibrium angular velocity.  It was also found that the Hall effect
is important only when the ion cyclotron frequency is comparable to,
or smaller than, the orbital frequency.  This situation may occur in
early galaxy formation where the magnetic fields are still weak, or in
weakly ionized protostellar disks.  \cite{Krolik06} have suggested
that, in this limit, short-wavelength modes are likely suppressed by
viscous or resistive damping, leaving only slowly growing,
long-wavelength modes as the magnetic field get sufficiently weak.
However, their analysis is restricted to low-$\beta_i$ plasmas as they
do not consider FLR effects, which we show to be much more important
than the Hall effect in the weak-field limit.  The strong FLR
stabilization of the MRI in the weak-field limit may have important
implications for the possible role of the MRI in the amplification of
weak, primordial magnetic fields.

The gyroviscous stress is defined as the traceless, perpendicular part
of the ion stress tensor which does not depend explicitly on the
collision frequency \citep{Ramos05}.  In typical cases, this stress
arises primarily from variations in particle drift velocities over the
scale of a Larmor orbit \citep{Kaufman60}.  However, other effects may
contribute to this stress, including gradients in heat fluxes.
Braginskii's form of the ion gyroviscous stress is appropriate for
collisional plasmas (in the sense that the ion mean-free-path is small
compared to the hydrodynamic perturbation length-scale), and in the
limit where the ion cyclotron period is short compared to collisional
and hydrodynamic time-scales \citep{Braginskii65}.  More general, but
more complicated, expressions for the gyroviscous force have been
derived, which are applicable to a broader range of collisionality
regimes and dynamical time scales; \cite{Ramos05} and references
therein provide derivations and discussions of these alternate forms.
We choose to work with Braginskii's form here because it is
(relatively) simple and applies to a broad range of astrophysical
objects.

The MRI in the collisionless regime, where the collisional
mean-free-path is greater than the mode wavelength, has been explored
by \cite{Quataert02} and \cite{Sharma03} using kinetic closures.
\cite{Islam05} have extended the single-fluid MHD treatment to lower
collisionality regimes by including the \cite{Braginskii65} form of
the parallel viscosity, and have obtained results similar to those
obtained using kinetic closure.  These various analyses have found the
linear growth rate to be enhanced by a factor of order unity at lower
collisionality when an azimuthal component of the magnetic field is
present, but the criterion for instability was found not to differ
from the MHD result.  We do include the parallel viscous stress for
completeness in our analysis, as formally it may be larger than the
gyroviscous stress. However, for the sake of simplicity we restrict
the MRI mode wavevector and magnetic field to be normal to the
accretion disk, in which case the parallel viscosity has no effect on
the MRI.  This case is the most unstable one in the collisional limit,
which is the limit in which we are mainly interested.

\section{Linear Theory}

\subsection{Model}

We consider the two-fluid MHD equations:
\begin{mathletters}
  \label{eq:model}
\begin{eqnarray}
  \frac{\partial n}{\partial t} & = & -\div{(n \v)} 
  \\ 
  n \frac{\partial \v}{\partial t} & = & -n \v \cdot \grad{\v} + 
  \frac{\J \times \B}{c}  - \mbox{} \\
  && \mbox{} - \grad{p} - \div{\tensor{\Pi}} - n g(r) \hat{r}
  \\
  \frac{\partial \B}{\partial t} & = & -c \curl{\E}
\end{eqnarray}
\end{mathletters}
where
\begin{eqnarray*}
  & \E = -\frac{1}{c} \v \times \B + \frac{1}{n e} 
  (\frac{1}{c}\J \times \B - \nabla p_e) 
  \\ 
  & \J = \frac{c}{4\pi} \curl{\B};\quad p = p_i + p_e. &
\end{eqnarray*}
The terms representing two-fluid effects are the Hall term ($\J \times
\B/n e c$) and the electron pressure gradient in the definition of $\E$.
We assume barytropic pressure variations of the form
\begin{equation}
  dp_s = \Gamma T_s dn
\end{equation}
for each species $s$ where, for example, $\Gamma=5/3$ for an adiabatic
equation of state.  For $\tensor{\Pi}$ we use the leading order terms
in the \cite{Braginskii65} closure:
\begin{displaymath}
  \tensor{\Pi} = \tensor{\Pi}^{v} + \tensor{\Pi}^{gv},
\end{displaymath}
where
\begin{eqnarray}
  \tensor{\Pi}^{v} & = & \eta_0 \frac{p_i}{2 \nu_i} 
    \left( \tensor{I} - 3 \b \b \right)
    \left( \b\cdot\W\cdot\b \right)
  \\
  \label{eq:gyroviscosity}
  \tensor{\Pi}^{gv} & = & \frac{p_i}{4 \omega_{c i}} \left\{
    \b \times \W \cdot (\tensor{I} + 3 \b\b) + \mbox{} \right. \\ 
    & & \left. \mbox{} + 
    \left[\b \times \W \cdot (\tensor{I} + 3 \b\b)\right]^\top \right\}
\end{eqnarray}
where $\b = \B/B$, $\omega_{c i} = e B_{z 0}/m_i c$ is the ion
cyclotron frequency, $\nu_i$ is the ion collision frequency, and the
coefficient $\eta_0 \approx 0.96$ is a factor determined by kinetic
analysis \citep{Braginskii65}.  The rate-of-strain tensor is
\begin{displaymath}
  \W = \nabla \v + (\nabla \v)^\top - \frac{2}{3} \tensor{I} \div{\v}.
\end{displaymath}
$\tensor{\Pi}^{v}$ is the parallel viscosity term considered by
\cite{Balbus04} and \cite{Islam05}, which represents the lowest-order
in $\partial_t/\nu_i$ correction to the fluid equations.  (It will be
shown, however, that the parallel viscosity plays no role in the MRI
in the configuration we choose to examine.)  $\tensor{\Pi}^{gv}$ is
the gyroviscous force, which represents the lowest-order FLR
correction to the fluid equations.  This form of $\tensor{\Pi}$ is
appropriate in the limit where $\omega_{c i} \gg \nu_i$.  Together
with the short mean-free-path condition ($k_\parallel \lambda_{mfp}
\ll 1$, where $\lambda_{mfp} \sim v_t / \nu_i$) necessary to justify
the fluid closure, this restricts the validity of our results to the
parameter regime where
\begin{equation}
\label{eq:validity}
k_\parallel v_t \ll \nu_i \ll \omega_{c i}.
\end{equation}
In the case of an accretion disk where $k_\parallel \agt 1/d$ is
limited by the disk height $d \sim v_t/\Omega$, this validity
condition becomes $\Omega \ll \nu_i \ll \omega_{c i}$.

\subsection{Equilibrium}

We consider a cylindrical equilibrium, using the standard coordinates
$(r,\theta,z)$, and assume axisymmetry in the azimuthal coordinate
($\partial_\theta = 0$).  For simplicity, we assume the equilibrium
magnetic field is of the form $\B_0 = B_{z 0} \hat{z}$, and that the
equilibrium fluid velocity is of the form $\v_0 = r \Omega(r)
\hat{\theta}$.  Without loss of generality, we orient our coordinate
system so that $\Omega(r) > 0$ at the radius of interest.  For such a
configuration, radial force balance is satisfied when
\begin{equation}
  \label{eq:radial_force_balance}
  g(r) = r \Omega^2(r) +
  \frac{\rho_i^2 \omega_{c i}}{2 r^2}\frac{\partial}{\partial r}
  \left[r^3 \Omega'(r)\right],
\end{equation}
where $\rho_i = v_t/\omega_{c i}$ is the ion Larmor radius, and $v_t =
\sqrt{T_i/m_i}$ is the ion thermal velocity.  We neglect any
equilibrium structure in the $z$-direction.  In the following
derivation, we choose to use equation~(\ref{eq:radial_force_balance})
to eliminate $g(r)$ in favor of $\Omega(r)$ (hereafter we will drop
the explicit dependence of $\Omega$ on $r$).

\subsection{Local Linear Dispersion Relation}

\begin{figure*}[t]
  \includegraphics[width=2.8in]{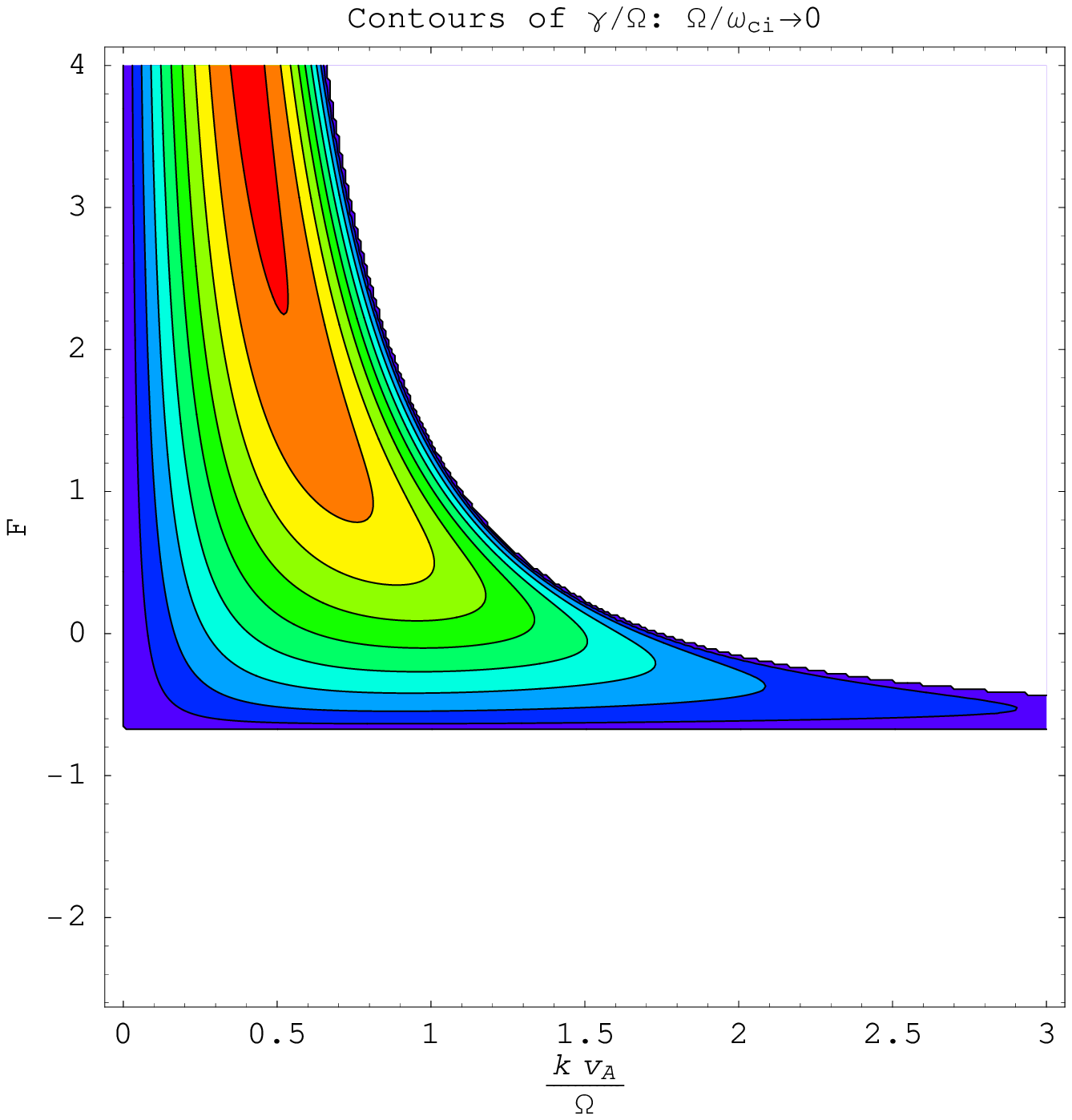}
  \includegraphics[width=2.8in]{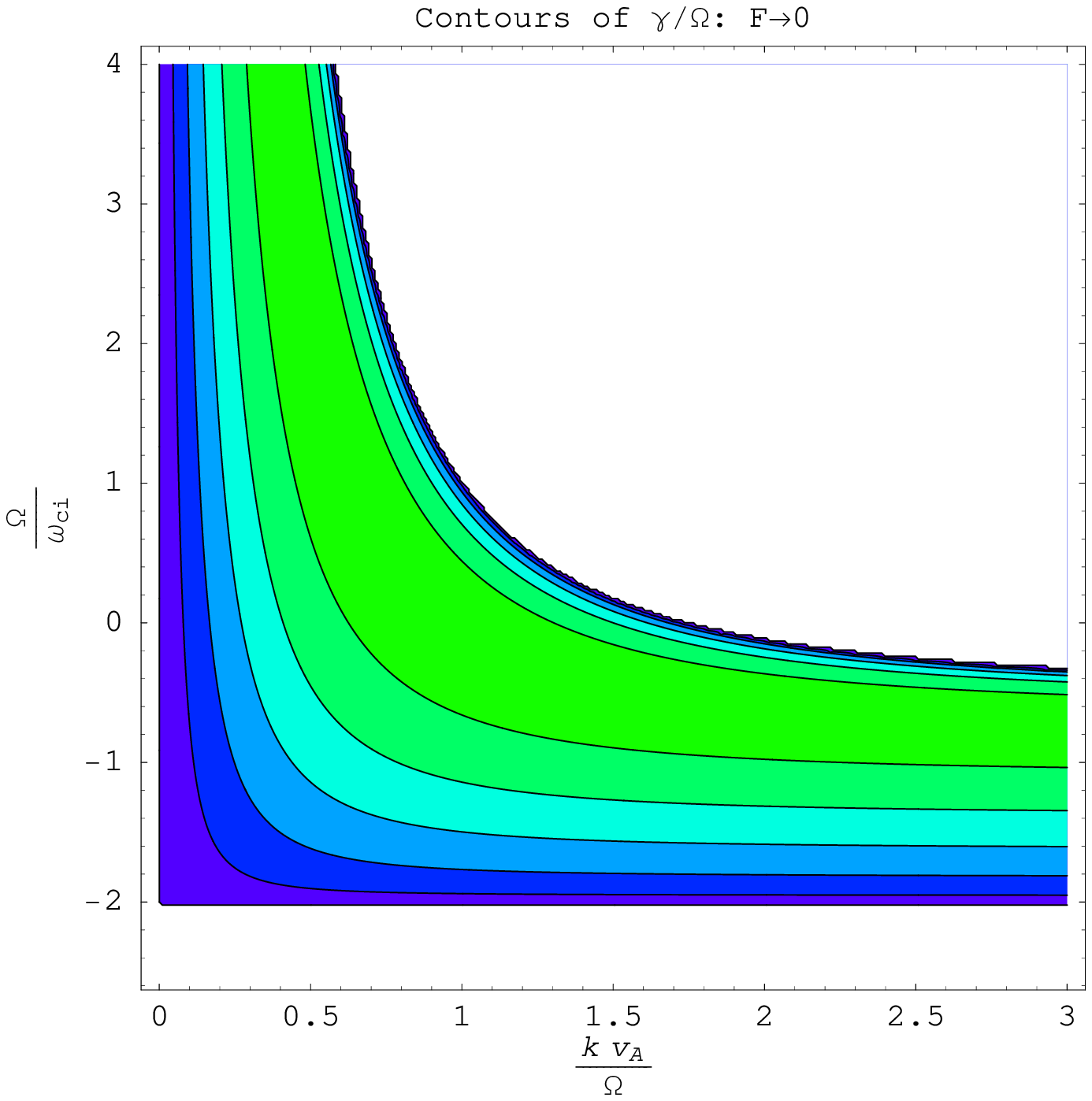}
  \includegraphics[width=0.6in]{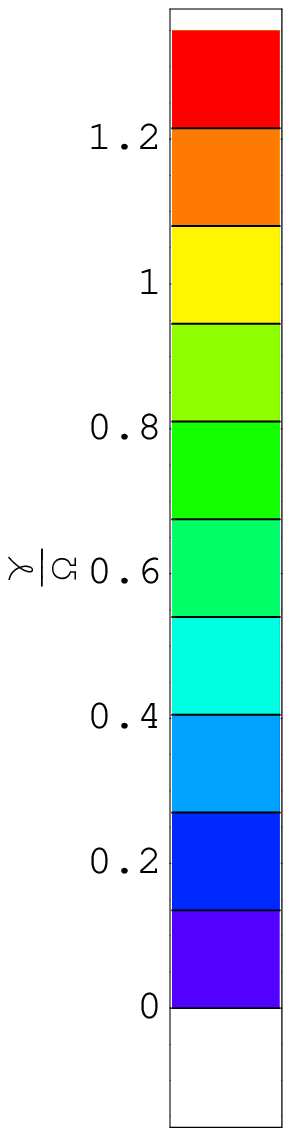}
  \caption{\label{fig:MRI_gamma} Contours of the growth rate of the
  unstable solution to equation~(\ref{eq:dispersion_relation}) are
  plotted versus $k v_A/\Omega$.  \emph{Left:} The growth rate is
  plotted for various values of $F$, which measures the importance of
  FLR effects, in the limit where $\Omega/\omega_{c i} \ll 1$ (no Hall
  effect).  $F = 0$ is the ideal MHD result.  \emph{Right:} The growth
  rate is plotted for various values of $\Omega/\omega_{c i}$, which
  measures the importance of the Hall effect, in the limit where $F
  \ll 1$ (no FLR effects).  $\Omega/\omega_{c i}=0$ is the ideal MHD
  result.}
\end{figure*}

We consider linear perturbations about this equilibrium having scale
lengths $1/k$ much smaller than the equilibrium flow gradient scale
length $L = \Omega/\Omega'$, so that $\delta = 1/ |k L| \ll 1$.  In
this limit, we may assume that the normal modes of the system are
plane waves to lowest order in $\delta$.  We restrict our analysis to
perturbations $\propto e^{i(k z - \omega t)}$ for simplicity, as this
is the most unstable case in both the MHD and collisionless limits
\citep{Quataert02}.  Carrying out the linearization of
equations~(\ref{eq:model}) yields the following dispersion relation,
to lowest order in $\delta$:
\begin{eqnarray}
  0 & = & (W^2 + W \frac{8 i \eta_0}{3} \frac{\omega_{c i}}{\nu_i}A - C)
  \times \mbox{} \nonumber \\ & & \mbox{} \times
  \left\{ W^4 + a_2 W^2 + a_0 \right\} \label{eq:dispersion_relation}
\end{eqnarray}
where
\begin{eqnarray*}
  a_2 & = & -2(2+R+K) + \mbox{} \\
  && \mbox{} + 2 A (4+R-2A) - H(R+H)
  \\
  a_0 & = & \left[K - A(R+2H) + 2(R+H) \right] \times \mbox{} \\
   && \mbox{} \times \left[K - A(R+2H) + H(2+R) \right].
\end{eqnarray*}
We define the dimensionless quantities
\begin{mathletters}
\begin{eqnarray}
  W &=& \omega/\Omega\\
  K &=& (v_A k/\Omega)^2\\
  A &=& K \beta_i (\Omega/\omega_{c i})/4 \equiv K F \label{eq:def_A}\\
  H &=& K (\Omega/\omega_{c i})\\
  C &=& (c_s k/\Omega)^2\\
  R &=& r \Omega'/\Omega
\end{eqnarray}
\end{mathletters}
and characteristic velocities $v_A^2 = B_{z 0}^2/4\pi n_0 m_i$ and
$c_s^2 = \Gamma (T_e + T_i)/m_i$.  Here $\beta_i = 8 \pi p_{i
0}/B_0^2$ is the ratio of ion thermal pressure to magnetic pressure.
The dimensionless parameter $A$ measures the importance of the
gyroviscous force, and setting $A = 0$ is equivalent to omitting $\div
\tensor{\Pi}^{gv}$ in the ion force equation.  Similarly, $H$ measures
the importance of the Hall term in Ohm's law, and $K$ measures the
importance of magnetic tension.  $R$ is the ratio of the radial
coordinate to the equilibrium flow gradient scale length, and is taken
to be $\sim \mathcal{O}(1)$.  For a Keplerian disk, $R = -3/2$.  For
convenience we have also defined
\begin{displaymath} 
  F = \frac{\beta_i}{4} \frac{\Omega}{\omega_{c i}},
\end{displaymath}
which is the ratio of the gyroviscous force to the magnetic tension
force.

Equation~(\ref{eq:dispersion_relation}) contains two uncoupled modes.
The first factor contains the acoustic mode, which may be damped by
the parallel viscosity when $A \ne 0$, and is not of interest here.
(While $\omega_{c i}/\nu_i$ is generally large, $A \omega_{c i}/\nu_i
\sim C \Omega/\nu_i \ll C$ in the collisional regime, so the effect on
the acoustic mode is small.)  The second factor, enclosed in braces,
contains the MRI.  In the limit where $A \to 0$ (no FLR effects), $H
\to 0$ (no Hall effect), and $\Omega/\nu_i \to 0$ (collisional
regime), the dispersion relation of \cite{Balbus98} is recovered.
Note that the parallel viscosity ($\propto \eta_0$) affects only the
acoustic mode and not the MRI.  Evidently, there is no
$\mathcal{O}(\Omega/\nu_i)$ correction to the MHD result for the MRI
when $\B_0 = B_{z 0} \z$, which is in agreement with the findings of
\cite{Sharma03} and \cite{Islam05}.  Here we are interested in
corrections to the collisional mode, for which the $\B_0 = B_{z 0} \z$
case is the most unstable.  Extending this analysis to a more general
magnetic field configuration substantially complicates the analysis.

It should also be noted that $\omega_{c i}$ is a signed quantity since
it is proportional to $B_z$, which may be positive or negative.  Since
we have chosen the coordinate system so that $\Omega$ is positive,
$\text{sign }\omega_{c i} = \text{sign }\B \cdot \vec{\Omega} =
\text{sign } F$.  It has been shown previously that the effect of the
Hall term on the MRI depends strongly on the sign relative signs of
$\omega_{c i}$ and $\Omega$ \citep{Wardle99}.  The effect of the
gyroviscous force has a similar dependence.

The growth rate $\gamma = \text{Im }\omega$ of the unstable solution
to equation~(\ref{eq:dispersion_relation}) is plotted in
figure~\ref{fig:MRI_gamma} for a Keplerian rotation profile
($R=-3/2$).  Note that the abscissa should be read as a normalized
wavenumber and not a normalized magnetic field strength, because $F$
and $\Omega/\omega_{c i}$ are dependent on $B$.  When $\omega_{c i} >
0$, and hence $F>0$ also, both the FLR and the Hall effects can be
seen to move the most unstable mode to lower wavenumbers, and to
reduce the value of $K$ at which the MRI is completely stabilized.
Also, FLR effects increase the growth rate of the most unstable mode.
When $\omega_{c i} < 0$, and hence $F<0$, both effects are seen
instead to increase the cutoff value of $K$ all the way to the point
where modes of any wavelength for which this analysis is valid are
unstable.  Presumably, the inclusion of a finite resistivity would
damp this resonance, as it does in the case of the resonance due to
the Hall effect \citep{Balbus01}.  When $F$ or $\Omega/\omega_{c i}$
becomes sufficiently negative ($F < -2/3$ when $\Omega/\omega_{c i}
\to 0$, or $\Omega/\omega_{c i} < -2$ when $F \to 0$), all values of
$k$ are suddenly completely stabilized.  (It has been shown by
\cite{Balbus01} that this stabilization is less sudden when finite
resistivity is included.)

\subsection{Instability Criterion}

Applying the Routh-Hurwitz theorem to
equation~(\ref{eq:dispersion_relation}), we find that the condition
for stability of an MRI mode is that:
\begin{mathletters}
\label{eq:stability_criterion}
\begin{eqnarray*}
  a_2 < 0 \mbox{ and } a_0 > 0.
\end{eqnarray*}
\end{mathletters}
This criterion is highly complicated, and for general values of $A$
and $H$, there may be multiple stable and unstable regions in
$K$-space.

In the ideal limit, when $A \to 0$ and $H \to 0$, the instability
criterion of \cite{Balbus98}, $K < -2 R$, is recovered.  This limit is
well understood, and in this case stabilization at high-$K$ is due to
the effect of magnetic tension.  In this limit, instability does not
exist in flows in which the angular velocity increases with radius
($R>0$).

The limit $A \to 0$, in which the Hall effect is dominant over the FLR
effects, has also been considered before.  In this limit,
equation~(\ref{eq:dispersion_relation}) reduces to the dispersion
relation of \cite{Balbus01}.  Since $A/H = \beta_i/4$, this limit
describes accretion disks having $\beta_i \ll 1$.  Formally, the
instability criterion in this case remains somewhat complicated
because the signs and relative magnitudes of most of the terms are
undetermined in general.  There is some discussion of the instability
criterion in this case by \cite{Wardle99} and \cite{Balbus01}, as well
as insight into its physical meaning.  We will not repeat this
discussion, except to mention a few interesting points.  The first is
that there may exist some values of $\Omega/\omega_{c i}$ for which
modes of any wavelength are unstable (this is true in the Keplerian
case for $-2 < \Omega/\omega_{c i} < -1/2$).  Also, some unstable
modes may be present in disks in which angular velocity increases with
radius ($R > 0$), in contrast to the ideal result \citep{Balbus01}.

We are more interested in the opposite limit, $\beta_i \gg 1$, in
which FLR effects are dominant over the Hall effect.  Taking $H \to
0$, the dispersion relation for the MRI reduces to
\begin{eqnarray}
\label{eq:flr_dispersion_relation}
0 & = & W^4 - \mbox{} \\
 && \mbox{} - 2 \left[ 2+R+K - A (4+R-2A) \right] W^2 + \mbox{} \nonumber \\
 && \mbox{} + \left(K- A R + 2 R \right) \left(K - A R \right) \nonumber
\end{eqnarray}
and the criterion for instability is found to be
\begin{equation}
  K  < -\frac{2 R}{1 - F R}.
\end{equation}
For the usual case where $R < 0$, all modes are completely stabilized
if $F < -1/|R|$.  As with the Hall effect, gyroviscosity allows
unstable modes to exist when $R > 0$; in this case, unstable modes may
exist when $F>1/R$.  In the limit where the gyroviscous force
dominates the force of magnetic tension, $F \gg 1$, the instability
criterion becomes simply
\begin{equation}
  \label{eq:FLR_instability_criterion}
  A < 2 \mbox{ and } F > 0
\end{equation}
or, equivalently, $(k \rho_i)^2 < 4 \Omega/\omega_{c i}$, In the case
where $F<0$, there are no unstable modes in this limit.

\section{Discussion and Conclusions}

The gyroviscous stress arises from changes to the guiding-center
drifts due to the gradients of magnetohydrodynamic forces (the
electric field, in this case) across a gyro-orbit.  A more complete
and quantitative discussion of this effect is discussed by
\cite{Kaufman60}.  Due to their larger Larmor radius, the ions are
more affected by this modification that the electrons, leading to the
generation of currents.  Essentially, the stabilization (or
destabilization) of the MRI by FLR effects is due to the currents
generated by the gyroviscous force being out of phase (or in phase)
with the current of the ideal-MHD MRI eigenmode.

The effect of the gyroviscous stress becomes relatively more important
to the behavior of the MRI as the magnetic field strength is
decreased.  The relative importance of this stress to the magnetic
tension is $A/K = F \propto B^{-3}$, and the relative importance of
the gyroviscous stress to the Hall effect is $A/H = \beta_i/4 \propto
B^{-2}$.  Note that this means analyses of the MRI in the
``weak-field'' limit (in the sense that $\Omega/\omega_{c i} \ga 1$)
which did not consider FLR effects are valid only for $\beta_i \ll 1$.
If the magnetic field is sufficiently weak, the validity condition for
the Braginskii closure, equation~(\ref{eq:validity}), may be violated
since $k v_t / \omega_{c i} = \sqrt{2 H F} \propto B^{-1}$, in which
case the above analysis no longer strictly applies.  Of course,
gyroviscosity does not shut off at this point; indeed, FLR effects are
expected to become increasingly powerful as the magnetic field is
decreased further, though not in a way that is correctly described by
equation~({\ref{eq:gyroviscosity}).  Therefore it is probable that the
low-$K$ modes which survive at the lowest magnetic field strengths
covered by this analysis will be completely stabilized as the magnetic
field decreases further.  This is the proper resolution to the
inconsistency of the ideal MHD result than the MRI remains unstable as
$B \to 0$ in the non-dissipative case.  Because the MRI is not present
at arbitrarily low magnetic fields, its role in the amplification of
primordial astrophysical magnetic fields is severely restricted.

It should also be noted that FLR effects may completely stabilize MRI
modes having wavelengths much greater than the ion Larmor radius,
where $k \rho_i \ll 1$.  As was shown by \cite{Rosenbluth62}, this is
possible for ``weakly unstable'' modes like the MRI.  Restricting our
analysis to the FLR-dominated limit ($F \gg 1$), MRI modes are
stabilized when $(k \rho_i)^2 > 4 (\Omega/\omega_{c i})$, where
$\Omega/\omega_{c i}$ is typically small in astrophysical accretion
disks.  This stabilization may be made more important by the fact that
in an accretion disk the lower bound on $k$ is set by the height of the
disk $d$, which may be much smaller than the equilibrium flow gradient
scale length $L$.  Although a proper understanding of this phenomenon
must take into account the $z$-stratification of the disk equilibrium,
we may estimate that the smallest wavenumber present in the disk is
$\sim \pi/d$.  The criterion for complete stabilization by FLR effects
of \emph{all} MRI modes within an accretion disk at some distance from
the central mass is then
\begin{equation}
  \pi^2 (\rho_i / d)^2 \agt 4 \Omega/\omega_{c i}.
\end{equation}
In the typical case where $d \sim v_t/\Omega$ this inequality reduces
to $\omega_{c i}/\Omega \alt \pi^2/4$.  While this is not typically
satisfied in astrophysical accretion disks, it may be satisfied in
nascent galaxies with weak magnetic fields, or weakly ionized
protostellar disks \citep{Krolik06}.  It is important to recall here
that the condition for validity of our analysis for wavelengths of
this scale requires $\Omega \ll \omega_{c i}$; whether FLR effects are
more or less stabilizing than our result would predict outside this
range of validity is a matter for further research.

\begin{figure}[t]
  \begin{center}
    \includegraphics[width=3in]{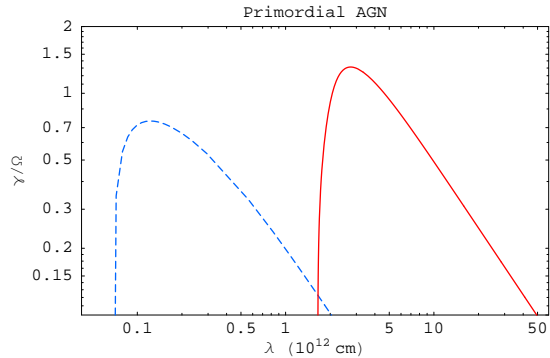}
  \end{center}
  \caption{\label{fig:galactic_MRI} The growth rate of the MRI is
    plotted versus wavelength for a Keplerian, $10^7 M_\sun$ active
    galactic nucleus in a weak magnetic field ($B = 10$ nG), at a
    distance $10^{16}$ cm.  The solutions to the dispersion relation
    including (solid line), and excluding (broken line) the FLR
    correction are plotted.  Gyroviscosity completely stabilizes the
    shorter-wavelength modes, and enhances the longer-wavelength
    modes.}
\end{figure}

For a concreteness, we consider the example of an active galactic
nucleus (AGN) in a weak magnetic field, for which the conditions for
validity of the Braginskii equations are met, and in which the effect
of the gyroviscous force is particularly strong.  We assume Keplerian
rotation ($R = -3/2$), a central mass of $10^7 M_\sun$, $B=10$ nG, $n
= 10$ cm$^{-3}$, and $T_i = 1$ eV, at a distance of $10^{16}$ cm.  For
this case, the relevant dimensionless parameters are $F \approx 384$,
$H \approx 3.81\times10^{-4}$, and $K \approx 1.96 \times 10^{-5}$ at
$k = \Omega/v_t$.  The growth rates of the MRI under these conditions
are plotted in figure~\ref{fig:galactic_MRI}.  From that figure it can
be seen that in both cases the FLR correction increases the wavelength
of the fastest-growing mode from $\approx 10^{11}$ cm to $\approx
25\times 10^{11}$ cm, and similarly increases the cutoff wavelength
below which there are no unstable modes.  In this example, the
weakness of the magnetic field is crucial to the importance of the FLR
effect.  If the magnetic field strength is raised by two orders of
magnitude, the FLR correction becomes insignificant.  Therefore,
``primordial'' accretion disks in weak magnetic fields will be most
affected by the FLR correction, whereas disks in stronger fields ($B
\ge 10$ $\mu$G) may be completely unaffected, unless they are
unusually hot and dense, or have an orbital frequency not much smaller
than the ion cyclotron frequency.

The nonlinear saturation of the MRI cannot be properly addressed by
linear analysis.  Because gyroviscosity is not dissipative, and there
is no energy associated with the gyroviscous term, one might expect
that the ultimate nonlinear saturated state should not be affected
strongly by the gyroviscosity.  Ultimately, questions of nonlinear
saturation should be addressed by numerical simulation.

\acknowledgments{The author is grateful for helpful discussions with
  Dr. Gregory Hammett and Ian Parrish.}


\end{document}